\documentclass[twoside,fleqn]{article}
\usepackage{amssymb}
\usepackage{npb}
\usepackage{graphics}
\newcommand{\be}{\begin{equation}}
\newcommand{\ee}{\end{equation}}

\def\beq{\begin{eqnarray}}    
\def\eeq{\end{eqnarray}}      

\def\be{\beta}

\def\La{\Lambda}

\def\si{\sigma}

\def\La{\Lambda}

\title{An Overview of the Anomaly-Induced Inflation}

\author{Ilya L. Shapiro 
\address{Departamento de Fisica, ICE, Universidade Federal de Juiz de
Fora, MG, Brazil}}
 
\begin{document}

\begin{abstract}
The anomaly-induced inflation (modified Starobinsky model)
is based on the application of the effective quantum field 
theory approach to the Early Universe. We present a brief 
general review of this model with a special attention to 
the existing difficulties and unsolved problems.
\begin{flushright}
\end{flushright}
\end{abstract}

\maketitle


$\,\,\,\,\,\,\,$
The original version of the anomaly-induced inflation 
\cite{fhh,star,vile,ander} is the cosmological model which 
takes into account the vacuum quantum effects of the free, 
massless and conformally coupled to metric matter fields
\cite{birdav}. The quantum correction to the Einstein 
equation with cosmological constant
\beq
R_{\mu\nu}\,-\,\frac12\,R\,g_{\mu\nu}\,=\,
8\pi G\,<T_{\mu\nu}>\,-\,\La
\label{1}
\eeq
produces a non-trivial effect because the anomalous trace
of the stress tensor 
\beq
T\,\,=\,<T_\mu^\mu>\,=\,
- \,(wC^2 + bE + c{\nabla^2} R)
\label{main equation}
\eeq
is non-zero. If the matter fields are absent, there are
the following equivalent ways to study the cosmological 
solution of (\ref{1}): using the $\,(0$-$0)$-component 
\cite{fhh,star} or via the anomaly-induced effective 
action \cite{buodsh,book}. Indeed, the last option is 
completely equivalent to taking the trace of (\ref{1}). 
The resulting equation has, for $\,k=0$ FRW metric, 
the following form (since the cases $k=\pm 1$ are 
quite similar \cite{asta} we will not consider them 
here):
\beq  
\frac{{\stackrel{....}{a}}}{a}
+\frac{{3\stackrel{.}{a}} {\stackrel{...}{a}}}{a^2}
+\frac{{\stackrel{..}{a}}^{2}}{a^{2}}
-\left( 5+\frac{4b}{c}\right) 
\frac{{\stackrel{..}{a}} {\stackrel{.}{a}}^{2}}{a^3}-
\\
-\frac{M_{P}^{2}}{8\pi c}
\left( \frac{{\stackrel{..}{a}}}{a}+
\frac{{\stackrel{.}{a}}^{2}}{a^{2}}
-\frac{2\Lambda }{3}\right)\,=\,0\,.
\label{foe}
\eeq
The equation above has a remarkable particular 
solution
\beq
a(t) \,=\, a_0 \cdot \exp(Ht)
\label{flat solution}
\eeq
where (motivated by the recent supernova data \cite{SN}, 
we consider only positive cosmological constant in the 
low-energy regime)
\beq
H\,=\, \frac{M_P}{\sqrt{-32\pi b}}\,\left(1\pm 
\sqrt{1+\frac{64\pi b}{3}\frac{\Lambda }{M_P^2}}\right)^{1/2}.
\label{H}
\eeq
As far as $\,\La \ll M_P^2$, we meet two very different 
solutions 
\beq
H_{c}\,=\,\sqrt{\frac{\Lambda }{3}}\,\,\,\,\,\,\,\,\,\,\,\,
{\rm and}\,\,\,\,\,\,\,\,\,\,\,\,
H_S\,=\,\frac{M_P}{\sqrt{-16\pi b}}\,.
\label{HH}
\eeq
The first solution is exactly the classical one, which  
one meets in the theory without quantum corrections, while 
the second one $\,H_S\,$ is the inflationary solution of 
Starobinsky. 

The phase portrait of the theory may 
look very different depending on the sign of the coefficient 
$\,c$ \cite{asta}. 
The inflationary solution is stable for a positive $\,c\,$ 
and is unstable in the case $\,c<0$. In the last case there 
are several stable points (attractors),
one of which corresponds to the FRW evolution. The original 
Starobinsky model deals only with the unstable solution. 
In this case one has to choose the initial conditions in a 
very special 
way. First of all, they must be very close to the exact 
exponential solution (\ref{flat solution}), such that
the inflation lasts long enough. Moreover the
choice of the initial condition has to provide that, after
the inflationary phase ends, the Universe will approach 
the attractor corresponding to the FRW solution, and not 
to the other (physically unacceptable) attractor. 
All the matter content 
of the Universe is created after the inflation ends through 
the decay of the massive degree of freedom induced by 
anomaly \cite{star,vile}. 

The Starobinsky model looks appealing, in particular 
because it is based on the quantum field theory results
and one does not need to introduce a special inflaton field. 
At the same time, this theory is somehow more difficult for 
investigation than the inflaton-based models,  
moreover it requires at least the same amount of the 
fine-tuning for initial conditions as many inflaton
models do. The importance of the very precise fine tuning
in the 3-dimensional space of initial data is due to the
existence of the non-FRW attractors corresponding to the 
physically
unacceptable behaviors \cite{wave}. These solutions resemble
the unphysical run-away solutions in QED and their 
possibility raised a 
generally suspicious relation to the idea of using quantum 
vacuum effects and higher derivative terms to achieve 
inflation. In order to advocate 
the whole approach, let us remark that the higher derivative
terms must be included into the gravitational action anyway,
because otherwise the quantum theory is inconsistent
(see, e.g. \cite{birdav,book}). Then there are only two ways 
to protect the theory from the undesirable effects of the 
higher derivatives. The first is what has been done in 
\cite{star} - choose the theory with the unstable inflation
$\,c<0\,$ and fine-tune the initial data. But there is a 
second way - to opt for a positive $\,c\,$ and stability 
of the exponential solution at the beginning of inflation.
The stable inflation is very robust with respect to the 
choice of the initial data and consequently it does not 
meet a problem with the unphysical solutions of the 
``run-away'' sort. 

The main problem of the stable inflationary model is to 
understand how the inflation ends. If we stay within the 
original framework \cite{fhh,star}, that is consider only 
massless conformal fields, the stable inflation will be 
eternal, obviously contradicting our experience. 
The modified Starobinsky model solves this problem
using the effective quantum field theory approach. The 
central idea has been suggested in \cite{insusy}, in the
framework of the physical interpretation of the stability
condition formulated before in \cite{wave}.

The coefficients $\,w$, $\,b\,$ and $\,c\,$ in 
(\ref{main equation}) depend on the number of the matter 
fields 
\beq
w =\frac{1}{360(4\pi)^2}
\left(
3{N_0} + 18{N_{1/2}} + 36{N_1}\right)\,,
\nonumber
\\
b =-\frac{1}{360(4\pi)^2}
\left(
{N_0} + 11{N_{1/2}} + 62{N_1}\right)\,,
\nonumber
\\ 
c =\frac{1}{360(4\pi)^2}
\left(
2{N_0} + 12{N_{1/2}} - 36{N_1}\right)\,.
\nonumber
\eeq
Using these relations the condition of stable inflation 
$\,c>0\,$ can be cast into the form
\beq
N_1 \,<\,\frac13\,N_{1/2}\,+\,\frac{1}{18}\,N_{0}\,.
\label{condition}
\eeq
The stable inflation requires the theory to include 
many scalars and fermions, for 
a given number of vectors. In high energy physics 
vectors correspond to the fundamental interactions. Then,
what is the physical interpretation of (\ref{condition})?
It is easy to see that the last inequality is not satisfied 
for the MSM with $\,N_{1,1/2,\,0}=(12,24,4)$. However, it 
is satisfied for the MSSM with $\,N_{1,1/2,\,0}=(12,32,104)$.
The same must be true for any realistic supersymmetric
model, because the supersymmetrization of the realistic 
model implies adding many fermion and scalar superpartners 
(sparticles) while the fundamental interactions are kept
the same. Then the transition between stable and unstable 
inflation can be associated with the SUSY breaking.
Let us remember that the SUSY breaking implies the special 
form of the mass spectrum, such that sparticles are 
very heavy compared to observable particles (otherwise
we should already see SUSY in the accelerator 
experiments). Therefore 
it is clear that inflation becomes unstable when its 
energy scale becomes smaller than the masses of the most 
of the sparticles and these sparticles decouple. For
definiteness, we shall associate the energy of inflation 
with the magnitude of the Hubble parameter $\,H$. 
Let us introduce the notation $\,M_*\,$ for the energy 
scale where the inequality (\ref{condition}) changes its 
sign to the opposite. The 
anomaly-induced inflation model assumes that the value of
$\,H\,$ is decreasing with time $\,{\dot H} < 0\,$ and that
$\,H_S\,$ is just an initial value of $\,H$. The stable 
inflation becomes unstable at the instant $\,t_f\,$ which 
is defined as a solution of the equation $\,H(t_f)=M_*$. 

Two recent relevant results of quantum field theory 
in curved space-time must be mentioned: 

\noindent
{\it i)} $\,$ 
The decoupling of the loops of massive fields in curved
space-time really takes place \cite{apco,fervi}. Despite 
serious difficulties in observing decoupling 
for the cosmological and inverse Newton constants, 
the existing data are sufficient for the 
purposes of the anomaly-induced inflation. In particular,
one can observe the smooth and monotone evolution of the
coefficient $\,c\,$ with scale and also the change of its
sign  from positive to negative due to the decoupling of 
the sparticles  \cite{apco}. 

\noindent
{\it ii)} $\,$ 
 The derivation of the effective action for massive
fields can be performed in an approximate but direct way 
using the conformal description of the massive fields
\cite{shocom}. This is achieved through the introduction 
of a new auxiliary scalar which enables one to establish 
a new conformal Noether identity in the classical theory.
At the quantum level, one meets conformal anomaly which 
can be integrated similar to the massless case 
\cite{shocom,asta}. The result of the whole procedure
is compatible with the renormalization group and may 
be viewed as its physical interpretation in the 
specific inflationary framework. Let us notice that the
general formulation of the renormalization group in 
curved space-time (see, e.g., \cite{book}) is based on 
the Minimal Subtraction scheme of renormalization and 
hence does not allow a physical interpretation for the 
specific cases. 

According to the method of \cite{shocom,asta}
the leading effect of the particle masses is that the 
Planck mass and the cosmological constant in the 
equation (\ref{foe}) and in the solution (\ref{H}) must 
be replaced by the variable expressions
\beq
M^2_P\,\to\,M^2_P\,(1-\tilde{f}\ln a)\,,
\label{central sigma}
\\
\La\,M^2_P\,\to\,\La\,M^2_P\,(1 - \tilde{g}\ln a)\,,
\label{central sigma C}
\eeq
where
\beq
\tilde{f} = 
\frac{1}{3\pi}\,\sum_{f}\,\frac{N_fm_f^2}{M_P^2} 
\,,
\\
\tilde{g} 
=\frac{1}{4\pi}\,\sum_{s}\frac{N_sm_s^4}{M_P^2\La}
\,-\,\frac{1}{\pi}\,\sum_{f}\frac{N_fm_f^4}{M_P^2\La}
\label{replace11}
\eeq
and the sums are taken over all species of fermions and
scalars with masses $\,m_f,\,m_s\,$ and multiplicities 
$\,N_f,\,N_s$. The numerical analysis shows that the 
unconstrained value of $\,\tilde{g}\,$ is difficult to 
interpret, hence for the sake of simplicity we shall 
suppose that the SUSY spectrum is constrained by the 
relation $\,\tilde{g}\cong 0$, and also assume a small 
value of $\,\La\,$ in the inflationary period. Then the 
equation (\ref{central sigma}) admits the following 
approximate analytical solution for $\ln a(t)=\si(t)$: 
\beq
\si(t)\,=\,H_0\,t\,-\,\frac{H^2_0}{4}\,\tilde{f}\,t^2\,.
\label{parabola}
\eeq
It is interesting that the numerical analysis confirms
the parabolic dependence (\ref{parabola}) with enormous
precision \cite{asta}. 

The relation (\ref{parabola})
can be used to evaluate the total number of the inflationary 
$\,e$-folds for different models of the SUSY breaking. 
The first option is MSSM with the value $M_*\sim 1\,TeV$.
It is easy to see that in this case 
$\,\tilde{f}\sim (M_*/M_P)^2 = 10^{-32}$ and 
therefore the total amount of the $\,e$-folds is $10^{32}$.
The expected temperature of the Universe after the end
of inflation can be evaluated from Einstein equation 
in a usual way $\,T \sim \sqrt{M_*\,M_P} = 10^{11}\,GeV$,
which is a standard estimate for the inflaton-based models.
An opposite extreme is to suppose that the SUSY takes place
only at a very high energies and is broken already at 
the GUT scale. Suppose $M_*\propto 10^{14}\,GeV$.
Then the total amount of $\,e$-folds is about $10^{10}$.
The problem of this version of the SUSY breaking is that 
the expected temperature after the end of inflation is very 
high $\,T \sim 10^{16}\,GeV$, such that inflation does not 
solve the monopole problem of GUT's. Hence, the 
anomaly-induced inflation really favors low-energy SUSY.

The problems of stability in the anomaly-induced inflation 
with respect to the perturbations of conformal factor and 
tensor degree of freedom will be given in the parallel 
presentations \cite{asta}. Let us consider 
some of the problems and potential difficulties of the 
anomaly-induced inflation.
\vskip 1mm

I. The stability of the inflationary solution 
(\ref{flat solution}) at the initial stage of inflation
is not absolutely safe, because the criterion $\,c>0\,$ 
concerns only the stability with respect to the 
perturbations of the conformal factor. We know that the 
metric has other degrees of freedom and the anomaly-induced 
effective action does not explain why the initial 
metric was homogeneous and isotropic. Even if the 
stability with the respect to the small tensor 
perturbations holds \cite{asta}, it does not solve the
problem completely, for the initial deviation from the
homogeneous and isotropic metric could be very large.
One can either 
suppose that these question may be answered within
a more fundamental theory like strings or try to find
the mechanism of the automatic isotropization of the
metric in the framework of the semiclassical theory.
The last possibility has been widely discussed
in the literature starting from \cite{zeld}, and 
indeed there are real chances to solve the problem in
the semiclassical effective framework.
\vskip 1mm

II. Let us 
make a very important observation concerning the 
coefficient $\,c\,$ in (\ref{main equation}). 
The value of $\,c\,$ can be modified by adding the 
$\,\int\sqrt{-g}R^2$-term 
to the classical action of vacuum. Furthermore, the 
different regularization schemes can produce, generally 
speaking, different results for this coefficient, and 
moreover the higher-loop effects in the theory with 
scalar fields produce the $\int\sqrt{-g}R^2$-type 
divergences, such that this term must be introduced 
from the very beginning at the classical level and its 
value must be fixed by the renormalization condition.
Hence it is unclear whether this indefiniteness may 
affect the modified Starobinsky model which is 
essentially based on the sign flip of the 
coefficient $\,c\,$ due to the possible SUSY breaking.
This problem has been recently analyzed in full details  
\cite{anomaio}. The conclusion relevant for us is that 
there is no ambiguity in the coefficient $\,c\,$ which can 
not be fixed by the renormalization condition. Hence, the
realization of the program of the anomaly-induced inflation
\cite{insusy} depends on the special renormalization 
condition for the $\,\int\sqrt{-g}R^2$-term. This 
renormalization condition must be imposed in such a way 
that the classical $\,\int\sqrt{-g}R^2$-term, together 
with the regularization ambiguity, would be small compared 
to the quantum contribution, e.g. derived in the 
point-splitting regularization \cite{chris} (see also 
discussion in \cite{hawk}). It is 
important to notice that this requirement is perfectly
compatible with the renormalization group equation.
\vskip 1mm

III. Let us finally comment on the relation between 
our model and the standard approach to inflation.
The inflaton models were introduced \cite{linde} in the 
situation when all possibilities to obtain inflation 
directly from particle physics did not look promising. 
From our point of view, these models represent a specific 
cosmological phenomenology (see, e.g. \cite{kolb}) which, 
as any other phenomenology, should serve as a bridge between 
observational data and some fundamental theory. In the
phenomenological setting the inflaton models have obvious
advantages compared to the anomaly-induced inflation. 
First of all, they are much more developed and also 
definitely simpler to deal with. However, the theoretical 
interpretation of an inflaton model is not going to be 
easy. It is supposed that one
can establish the form of the inflaton potential with some
reasonable precision and then find a quantum field theory
with a scalar field or fields,
string theory etc which, after all quantum corrections 
are taken into account, produces exactly the same potential.
The problem of establishing the form of potential may become
more difficult, technically, for the growing amount of the 
observational data, 
but the idea to establish a theoretical counterpart may be 
not realistic at all if the inflation was not caused by a 
scalar field potential.

On the other hand, the anomaly-induced inflation is also,
in part, phenomenological. At the present level of knowledge 
this model does not require a fine-tuning of initial data or 
other parameters, but it is improbable that this situation 
will persist when we start the analysis of the reheating 
and density perturbations.
Their behavior will likely depend on the high energy 
particle spectrum, on the details of decoupling for the 
inverse Newton constant and also on the choice of the 
vacuum for these perturbations which can not be performed, 
at the present state 
of art, in a unique way. In part, the same concerns 
the choice of vacuum for the metric perturbations 
at the beginning of inflation \cite{wave}. 

One has to remember that the inflaton models are
very useful as a reference point, because they always 
provide (maybe with more than one scalar field)
the fit with experimental or observational data with 
any desirable 
precision. In this way one obtains better understanding of 
these data and this is extremely important for the development 
of more complicated theory-based models such as the 
anomaly-induced inflation. Perhaps, the best solution would 
be to present the anomaly-induced effective action in a 
standard metric-inflaton form. An attempt of this 
sort has been undertaken in \cite{wave}. It turned out 
that one can perform such reduction, but only if one is 
exclusively interested in the evolution of the conformal 
factor. Within 
this approximation it is possible to present the effective 
action induced by quantum effects of matter field as a 
a second-derivative metric-scalar theory which can be 
of course called a new inflaton model. However, if we 
are interested, e.g., in the metric perturbations, one 
has to introduce more complicated auxiliary tensor 
``inflatons'' and the similarity with the usual 
inflationary models does not work.

All in all, despite the anomaly-induced inflation is 
not as developed as inflaton models, it represents an 
attractive alternative to them. In particular,
it enables one to avoid the standard fine-tuning in 
the choice of the initial data, gives a good 
chance to have a natural graceful exit and also to 
control the amplitude of the gravitational perturbations
\cite{asta}. Only further theoretical
and phenomenological study of this and other models and 
their comparison 
with future experimental/observational data may eventually 
show which of the models is closer to realities of our 
Universe in the first instants of its history.

\textit{Acknowledgments}: 
Author is grateful to CNPq for the fellowship and 
to FAPEMIG for the grant. 

\providecommand{\href}[2]{#2}

\end{document}